\documentclass[twocolumn,showpacs,amsmath,prl]{revtex4-1}

\usepackage{graphicx}
\usepackage{xcolor}

\definecolor{red}{rgb}{1.0,0.0,0.0}

\begin{document}

\title{Stacking-Fault Energy and Anti-Invar Effect in FeMn Alloys} 

\author{Andrei Reyes-Huamantinco}
 \email{a.huamantinco@mcl.at}
\affiliation{Chair of Atomistic Modeling and Design of Materials, University
of Leoben}
\affiliation{Materials Center Leoben Forschung GmbH, A-8700 Leoben, Austria}

\author{Peter Puschnig}
\affiliation{Chair of Atomistic Modeling and Design of Materials, University
of Leoben, A-8700 Leoben, Austria}

\author{Claudia Ambrosch-Draxl}
\affiliation{Chair of Atomistic Modeling and Design of Materials, University
of Leoben, A-8700 Leoben, Austria}

\author{Oleg E. Peil}
\affiliation{I. Institute for Theoretical Physics, University of Hamburg,
Germany}

\author{Andrei V. Ruban}
\affiliation{Applied Materials Physics, Royal Institute of Technology,
SE-10044 Stockholm, Sweden}
 
\date{\today} 
              
\begin{abstract} 

Based on state-of-the-art density-functional-theory methods we calculate
the stacking-fault energy of the paramagnetic random
Fe-22.5at.\%Mn alloy between 300--800~K. We estimate magnetic thermal
excitations by considering longitudinal spin-fluctuations.
Our results demonstrate that the interplay between the magnetic
excitations and the thermal lattice expansion is the main factor
determining the anti-Invar effect, the hcp--fcc transformation
temperature, and the stacking-fault energy,
which is in excellent agreement with measurements.

\end{abstract} 
 
\maketitle


The intrinsic stacking-fault (SF) is one of the simplest planar defects
of the crystal lattice. It is characterized by a fault in the usual $ABC$
stacking sequence of the fcc structure,  $...ABCAB|ABC...$, which resembles
locally the stacking sequence of the hcp structure.
The energy to create a stacking-fault, the stacking-fault energy (SFE),
is related to the ductility of the material, but the connection between the SFE
and the plastic deformation mechanism has so far only been recognized empirically,
for instance, in high-Mn steels~\cite{grassel00,frommeyer03}. Nevertheless,
the existence of such a connection provides a unique opportunity for
a first-principles theory to be a part of intelligent design of new materials,
and in particular high-performance steels, since the SFE
is an atomic-scale property readily accessible via \emph{ab initio} methods.

The \emph{ab initio} calculation of the SFE in real alloys at finite temperature
presents two main complications: First, the proper description of the atomic-scale
 structure of multicomponent alloys is complicated because 
the atomic configuration of alloy components is usually unknown, and,
moreover, it is still quite cumbersome to model. Second, the accurate
treatment of the thermal excitations of lattice vibrations and magnetic
fluctuations in alloys still remains a challenge for any theoretical
approach. The magnetic fluctuations are particularly important in fcc
and hcp Fe-based alloys, which are weakly itinerant magnets.
In particular, this concerns the technologically important Fe-(20-30)
at.\%Mn alloys, which exhibit both Invar and anti-Invar effects, as well
as martensitic transformations and shape memory
effects~\cite{witusiewicz04,schneider95}.

Special attention is given to the fcc Fe-22.5 at.\% Mn, which 
exhibits the lowest room-temperature SFE of the fcc Fe-Mn
alloys~\cite{volosevich76}, and thus is of particular interest for the
automotive industry. This alloy undergoes a structural transformation between
$\gamma$-austenite (fcc) and \mbox{$\epsilon$-martensite} (hcp) which,
on cooling, is characterized by the martensite start temperature,
$M_{s}$ = 375~K~\cite{cotes95} and, on heating, by the austenite start
temperature, $A_{s}$ = 450~K~\cite{cotes95}. 
Important is that the N\'eel temperatures of both fcc
(360~K~\cite{witusiewicz04}) and hcp (230~K~\cite{ohno71}) are below
$M_{s}$ and $A_{s}$, which implies that the fcc--hcp
transformation occurs in the paramagnetic state. 
In addition, the anti-Invar effect, or enhanced thermal lattice
expansion, that has been correlated to the magneto-volume
coupling~\cite{schneider95}, also takes place in the paramagnetic state.
It is, thus, clear that methods capable of treating the paramagnetic
state in a sufficiently reliable fashion are required for a proper
description of any finite-temperature properties in this system.

In this Letter, we study the SFE of the \mbox{Fe-22.5 at.\% Mn}
binary alloy between 300--800~K, using a simple but powerful model based on
density-functional theory to take into account spin fluctuations of Fe
and Mn atoms, which allows us to obtain excellent agreement with experimental
data~\cite{volosevich76}. The key feature of the methodology used is the
capability to describe the strong magneto-volume coupling at finite temperature.
We show then that the magneto-volume coupling is responsible for the
anti-Invar effect and the temperature dependence of the SFE in the
Fe-22.5at.\%Mn alloy.

Two recent \emph{ab initio} calculations of the SFE in FeMn alloys~\cite{kibey06,dick09},
that were done at zero-K, failed to reproduce the finite-temperature
experimental values even qualitatively. The main motivation of this work
is to show that the accurate treatment of the thermal excitations
of the magneto-volume coupling is essential for the calculation of the
temperature-dependent SFE for the fcc Fe-based alloys, in particular
Fe-22.5at.\%Mn.


The SFE is evaluated using the axial next-nearest neighbor Ising (ANNNI)
model~\cite{denteneer87}:
$SFE (T) = G^{hcp} (T) + 2G^{dhcp} (T) - 3G^{fcc} (T)$,
where $G(T)$ is the Helmholtz free energy,
and the volume per atom of the \emph{ideal} hcp and dhcp structures is the
same as for the fcc structure, implying that only the fcc volume needs
to be known. The alloy configuration is assumed to be completely random,
which is supported by recent experimental data~\cite{martinez09},
and the Mn concentration at the SF is the same as in the bulk,
i.e., the Suzuki effect~\cite{suzuki62} is inoperative due to the very
slow diffusion of manganese in the austenite phase~\cite{ueshima86}.
The ANNNI model is known to be less accurate than direct supercell
calculations, where the SF structure is treated explicitly and the
geometry can be relaxed~\cite{li11}. However, accurate SFE supercell
calculations for random alloys in the paramagnetic state are currently
impractical.

Generally, the free energy is a difficult quantity to evaluate.
Using the coarse graining of the partition function, the contributions
from electronic, magnetic, vibrational and configurational excitations
can be separated out based on their different time scales. 
Clearly, the entropy contribution from magnetic excitations in
the paramagnetic state can be significant and needs to be assessed.
As for the vibrational entropy, there are currently no available
theoretical tools to determine it accurately in paramagnetic random alloys. 
However, its contribution to the free energy \emph{differences} required
to evaluate the SFE has been argued to be small~\cite{vitos06}.
The configurational entropy does not contribute to the SFE in the case of
a completely random alloy. Lastly, we regard the effect of the local
lattice relaxations to be negligible since the atomic radii of iron
and manganese are similar.

Thus, the magnetic excitations seem to be the main entropy contributor
to the SFE. The relevant part of the Helmholtz free energy per site
is then determined as
$G(T) =  F_{el}(T) - T S_{mag}(T)$, where
$F_{el}(T) \equiv F_{el}\left[ S_{WS}(T),\{\bar{m}_{i}(T)\} \right] $
is the electronic part of the free energy, which includes one-electron
excitations and depends on the volume given in terms of the
Wigner-Seitz (WS) radius $S_{WS}(T)$ and the average local magnetic
moments $\bar{m}_{i}(T)$ of each alloy component $i$;
the magnetic entropy $S_{mag}(T) \equiv S_{mag}\left[\{\bar{m}_{i}(T)\}\right]$
is determined by the thermal fluctuations of the magnetic moments. 




We start with a detailed discussion of the equilibrium volume at finite
temperature $S_{WS}(T)$ and its relation to the magnetic moments
$\bar{m}_{i}(T)$. First, we consider the equation of state at $T=0$~K of
the fcc and hcp random alloys in the \emph{paramagnetic} configuration.
In order to provide an accurate account of the local environment effects
and energetics, we model the alloy by 384-atom supercells consisting of 298 Fe
and 86 Mn atoms distributed randomly on the underlying fcc and hcp lattices,
i.e., the alloy composition is Fe$_{0.776}$Mn$_{0.224}$.
The DFT self-consistent total energy calculations are done in the
disordered local moment (DLM)~\cite{gyorffy85} configuration by using the
locally self-consistent Green's function (LSGF)
method~\cite{abrikosov97, peil11} implemented within the exact muffin-tin
orbital method combined with the full-charge density technique
 (EMTO-FCD)~\cite{emto}, called LSGF-EMTO~\cite{lsgf_details}. 

\begin{figure}[tb] 
\begin{center} 
\includegraphics[width=\linewidth]{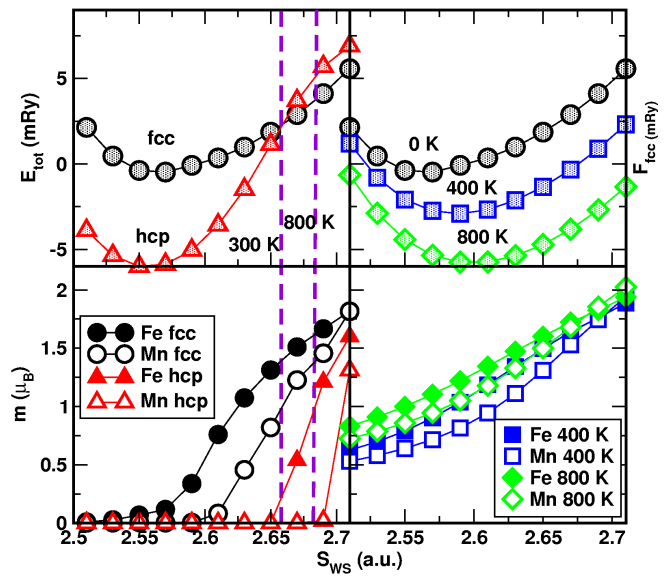} 
\end{center} 
\vspace{-5mm} 
\caption{(Color online) Total energies (upper-left panel) and zero-K DLM
magnetic moments (lower-left panel) of Fe and Mn in fcc
(circles) and hcp (triangles) Fe$_{0.776}$Mn$_{0.224}$, as function of volume.
Two dashed vertical lines show the \emph{experimental} WS radii for this
alloy at 300 and 800~K~\cite{li02,cotes95}. Helmholtz free energies
(upper-right panel) and finite-temperature DLM moments (lower-right panel)
of the fcc Fe$_{0.776}$Mn$_{0.224}$ alloy at 400~K (squares) and 800~K (diamonds).
The total energy of the Fe$_{0.776}$Mn$_{0.224}$ alloy at zero-K is shown
by filled circles.
} 
\label{fig:E_m} 
\end{figure} 

In the upper-left and lower-left panels of Fig.~\ref{fig:E_m}, we show, respectively,
the total energies and the DLM local magnetic moments of Fe and Mn as a function
of the WS radius for the fcc and hcp Fe$_{0.776}$Mn$_{0.224}$ alloys. 
We obtain a zero-K fcc equilibrium WS radius, 2.563~a.u., which is much
smaller than the room-temperature experimental value, 2.658~a.u.~\cite{li02},
indicated by the dashed vertical line in Fig.~\ref{fig:E_m}
~\cite{a_300K}. Our zero-K result is in agreement with a recent
full-potential supercell calculation, 2.574~a.u.~\cite{dick09}, while at
variance with a previous EMTO-FCD study, 2.605~a.u.~\cite{gebhardt10}.
The latter calculation, however, used the frozen-core approximation,
a smaller set of k-points, and had a freedom in the choice of unspecified
screening parameters, which can explain the disagreement.

From Fig.~\ref{fig:E_m}, we realize that both fcc and hcp phases possess
a low-spin equilibrium volume and exhibit a low-spin (LS)-- high-spin (HS)
transition with increasing WS radii, indicative of the anti-Invar
behavior~\cite{moruzzi90}.  In addition, the fcc structure becomes more
stable than hcp above 2.660~a.u., while simultaneously, a LS hcp--HS fcc
transition takes place, demonstrating the coupling between the structural
transformation and anti-Invar behavior in the paramagnetic
state, which has been discussed in the literature~\cite{moruzzi90}.


At finite temperature, the Helmholtz free energy must be calculated
instead of the total energy. The electronic single-particle excitations
are described via the Fermi function, while we use the simplest model 
to describe the longitudinal spin-fluctuations (LSF) in the DLM state.
Specifically, we prescribe to each site $p$ in the supercell,
having DLM magnetic moment $m_p$, the entropy $S_{mag} = k_{B} \ln(m_p + 1)$. 
The value of $m_p$ is determined by the minimization of the Helmholtz
free energy, and in this way, it corresponds to the average
magnitude of the magnetic excitations in the DLM state. 
We note that a more elaborate approach to the treatment of LSF will
be used below for the calculation of the SFE.


We have calculated the free energy as a function of the WS radius at
400 and 800~K using LSGF-EMTO, taking into account the LSF as described above. 
Compared with the zero-K result, the minima of the free energy curves
are significantly shifted towards larger volumes upper-right panel in
Fig.~\ref{fig:E_m}), which is caused by the drastic increase of the magnetic
moments due to the LSF (lower-right panel in Fig.~\ref{fig:E_m}).
From these data, one can estimate $S_{WS}(T)$ by taking into account
the vibrational entropy using the Debye-Gr\"uneisen (DG) model~\cite{moruzzi88}.
We obtain 2.650 and 2.698~a.u. at 400 and 800~K, respectively, in agreement with
the corresponding experimental $S_{WS}(T)$, 2.662 and 2.682~a.u.,
estimated from X-ray diffraction (XRD)~\cite{li02} and dilatometric
measurements~\cite{cotes95}.

We also obtain, however, a very large isotropic Gr\"uneisen constant,
about 4, and a strongly overestimated value of the thermal-expansion
coefficient, which shows that the DG model cannot be used for the
quantitative determination of $S_{WS}(T)$ in systems with a large
magnetically-induced anharmoncity. Nevertheless, the model
shows the strong magneto-volume coupling in this system and
allows us to capture qualitatively the crucial feature of the
anti-Invar effect, namely, enhanced thermal lattice expansion caused
by the so-called moment-volume instabilities~\cite{moruzzi90}, i.e.,
the thermally activated LS low-volume--HS high-volume transition.


The results show that the magneto-volume coupling is an essential feature
of the FeMn alloy, and it is, thus, important to have accurate values of
$S_{WS}(T)$, and the average local magnetic moment of each alloy component $i$, 
$\bar{m}_i(T)\equiv \bar{m}_{i}[T; S_{WS}(T)]$,
at a given temperature $T$, for a proper description of the system. 
There are no theoretical methods available to accurately calculate
$S_{WS}(T)$ for paramagnetic random alloys, but it can be measured.
The situation is, however, the opposite with respect to the magnetic
moments $\bar{m}_i(T)$ in the paramagnetic state, which are difficult
to measure,

but can be calculated by the procedure described below. 
We, therefore, combine the values of $S_{WS}(T)$ estimated from the
XRD~\cite{li02} and dilatometric measurements~\cite{cotes95} with
the theoretically evaluated values of $\bar{m}_{i}(T)$ to obtain
the free energy and finally the SFE at finite temperatures. 


To this end, we proceed to a more elaborate treatment of the
LSF and determine $\bar{m}_i(T)$ following the approach in Ref.~\cite{ruban07}
generalized to the case of a binary alloy. 
In this procedure, $\bar{m}_{i}(T)$ are evaluated as the thermodynamic
averages of the fluctuating magnetic moments of iron and
manganese at a corresponding temperature. For that purpose we assume
that the free energy is dominated by single-site magnetic fluctuations
in the paramagnetic DLM state. In this case, a Hamiltonian describing
the LSF can be defined as $H_{mag} = \sum_{i} J_{i}(m_{i}) c_{i}$,
with $i = $Fe, Mn, where $J_{i}(m_{i})$ is the energy necessary to excite
the magnetic moment $m_{i}$ of the $i$th alloy component. In calculating
 $J_{Fe(Mn)}$, we fix one of the magnetic moments, $m_{Fe(Mn)}$, while
letting another, $m_{Mn(Fe)}$, relax. 

\begin{figure}[tb] 
\begin{center} 
\includegraphics[width=\linewidth,angle=-90,scale=0.8]{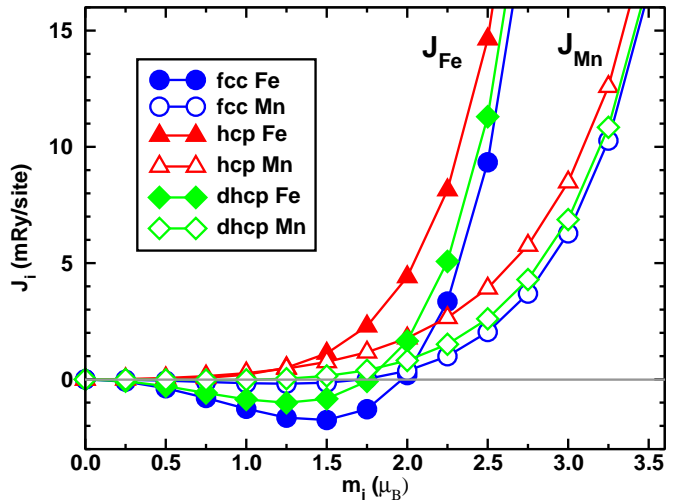} 
\end{center} 
\vspace{-5mm} 
\caption{(Color online) The energy of the LSF, $J_i$,
 obtained for different structures, at
 the fcc $S_{WS}(T = 360~K)$ = 2.660~a.u.~\cite{li02,cotes95},
 in the EMTO-FCD calculations.} 
\label{fig:J} 
\end{figure} 

In Fig. \ref{fig:J} we show the results for $J_{i}(m_{i})$ obtained for
fcc $S_{WS}(T = 360~K)$ = 2.660~a.u.~\cite{li02,cotes95}
in the EMTO-FCD calculations~\cite{cpa_details}.
In all the cases, $J_i$ have very shallow minima indicating low-energetic
accessibility of spin fluctuations, with large variety of the magnetic moments,
which reflects the weak itinerant nature of magnetism in this system.
The minima of the $J_i(m_{i})$ define the equilibrium magnetic moments in
the DLM ground state, $m^{DLM}_{i} \equiv \bar{m}_{i}(T = 0)$, and we obtain
$(m_{Fe}^{DLM},m_{Mn}^{DLM})$ = (1.4,1.1)$\mu_{B}$ for the fcc, 
(0.0,0.0) $\mu_{B}$ for the hcp, and (1.2,0.8) $\mu_{B}$ for the dhcp
structures, respectively.  Note that these $m^{DLM}_{i}$ are close to those
in the lower-left panel of Fig.~\ref{fig:E_m}, at $S_{WS}$ = 2.660~a.u.,
that have been obtained at zero-K in the LSGF-EMTO calculations.

\begin{figure} 
\begin{center} 
\includegraphics[width=\linewidth]{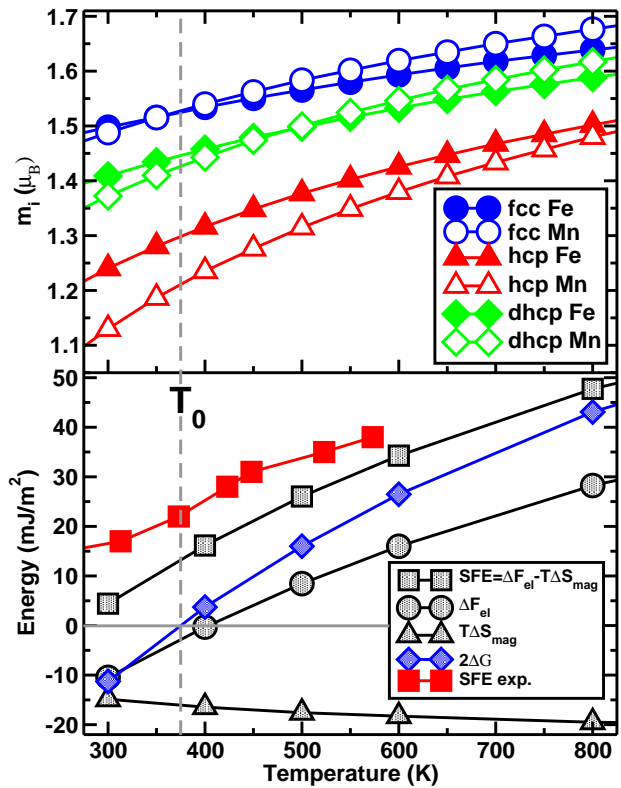} 
\end{center} 
\vspace{-5mm} 
\caption{(Color online) Average magnetic moments of Fe and Mn
in different structures (upper panel), and the SFE and its components
(lower panel) as a function of temperature. Vertical line indicates
the theoretical estimate of the hcp--fcc phase transition.} 
\label{fig:SFE_T} 
\end{figure}

At finite temperature $T$, the average magnetic moments, $\bar{m}_{i}(T)$, 
are readily obtained from the $J_i$ by a simple Monte Carlo integration
for each component. In doing so, we assume the complete coupling between
longitudinal and transverse fluctuations, i.e., the magnitude of the local
magnetic moment is determined by its $x$, $y$, and $z$-projections, when
integrating over the classical spin phase space.
The results are shown in the upper panel of Fig. \ref{fig:SFE_T} for different
structures in the temperature interval of 300--800~K,
and were obtained using the $J_i$ of Fig. \ref{fig:J}.
Note that
our room-temperature $\bar{m}_i(T)$ for the fcc phase is in agreement
with neutron diffraction room-temperature measurements of the average
moment per site in fcc Fe-30at.\%Mn, 1.5~$\mu_{B}$~\cite{umebayashi66}.


Finally, in order to obtain the SFE, we use the calculated
$\bar{m}_i(T)$ of Fig.~\ref{fig:SFE_T} together with the experimental
fcc $S_{WS}(T)$ in the free energy calculations of the
fcc, hcp and dhcp structures. In the lower panel of
Fig.~\ref{fig:SFE_T} one can readily see that despite a small quantitative
discrepancy with the experimental data~\cite{volosevich76}, the temperature
dependence of the SFE is nicely reproduced. Moreover, the entropic,
$T \Delta S_{mag}(T)$, and electronic free energy, $\Delta F_{el}(T)$,
contributions behave quite differently with temperature, with the latter
being fully responsible for the temperature dependence of the SFE.
The entropic contribution, on the other hand, just shifts (although 
quite substantially) the value of the SFE towards the experimental value.
This result is contrasted with earlier results by Vitos {\it et al.}
for FeCrNi alloys~\cite{vitos06}, where the temperature dependence of
the SFE was found to be determined solely by the magnetic excitations.

A detailed scrutiny of the sensitivity of the SFE to the values of
$S_{WS}(T)$ and $m_{i}(T)$ enables us to establish the crucial role
that thermal excitations and magneto-volume coupling play in the behavior
of the SFE. In particular, neglecting the temperature dependence of either
the volume, $S_{WS}$, or the magnetic moments results, respectively,
in a largely underestimated value of the SFE, -250 mJ/m$^{2}$ at 300~K,
when using the equilibrium theoretical zero-K $S_{WS}$ and $m_i(T)$ from
Fig.~\ref{fig:SFE_T}, or in an overestimated value of 150 mJ/m$^{2}$ at 300~K
with the experimental $S_{WS}(T)$ and zero-K $m^{DLM}_{i}$.

The obtained free energies allow us also to investigate the
structural stability by examining the so-called driving force
$\Delta G (T) = G^{hcp} (T) - G^{fcc} (T)$. Our value of the transition
temperature, determined by the condition $\Delta G(T_0) = 0$, is 375~K,
which is obtained using the ideal $c/a$ ratio and the same atomic volume
for the hcp phase as for the fcc one. It is in reasonable agreement with
the experimental estimate  $T^{\mathrm{exp}}_{0} \approx (M_{s} + A_{s})/2$
= 412~K~\cite{cotes95}.

We would like also to remark on the importance of including the dhcp
term to the expression for the SFE. In the axial nearest-neighbor Ising (ANNI)
model, the SFE is equal to $2 \Delta G$.
The comparison of the SFE calculated using the next-nearest-neighbor model
and $2 \Delta G$ in Fig.~\ref{fig:SFE_T} makes clear that the contribution
of the dhcp structure is important for the quantitative description of the SFE.


In summary, our results demonstrate that the interplay between the
LSF magnetic excitations and the thermal lattice expansion is the main factor
determining the anti-Invar effect, the hcp--fcc transformation temperature,
and the SFE in the paramagnetic random Fe-22.5at.\%Mn alloy. 
In principle, this strong sensitivity of the SFE with respect to
the quite elusive thermal effects of the magneto-volume coupling makes
theoretical studies in magnetic alloys rather difficult.
At the same time, we have clearly demonstrated that the accurate account
of important thermal contributions within the DFT \emph{ab initio}
approach allows not only to obtain a good quantitative agreement with
experiment but also to identify mechanisms dominating the formation and
stability of the SF at finite temperatures. The formalism presented here can
be further generalized to the case of austenitic stainless steels, which
are many-component systems with the presence of strong atomic short-range
order and local lattice relaxation effects, thereby opening a perspective
of the intelligent design of new high-performance materials.

\begin{acknowledgments}
A.R.H., P.P., and C.A.D. acknowledge the financial support by the COMET K2 Centre MPPE,
B\"ohler Special Steels, and Voestalpine~AG.
A.V.R. is grateful to the Swedish Research Council (VR), Hero-m centrum and the
ERC grant for financial support.
\end{acknowledgments}


\begin{thebibliography}{99}



\bibitem{grassel00} O. Grassel, L. Kruger, G. Frommeyer, and L. W. Meyer, Int. J.
Plasticity {\bf 16}, 1391 (2000).

\bibitem{frommeyer03} G. Frommeyer, U. Brux and P. Neumann, ISIJ Int. {\bf 43},
438 (2003).

\bibitem{witusiewicz04} V. T. Witusiewicz, F. Sommer and E. J. Mittemeijer,
J. Phase Equilibria Diffusion {\bf 4}, 346 (2004).

\bibitem{schneider95} T. Schneider, M. Acet, B. Rellinghaus, E. F. Wassermann,
and W. Pepperhoff, Phys. Rev. B {\bf 51}, 8917 (1995).

\bibitem{volosevich76} P. Y. Volosevich, V. N. Gridnev, and Y. N. Petrov,
Fiz. Metal. Metalloved. {\bf 42}, 372 (1976).

\bibitem{cotes95} S. Cotes, M. Sade, and A. Fernandez Guillermet,
Metallurgical Mat. Trans. A {\bf 26}, 1957 (1995). In Fig.~10, there
is an obvious typo in the scale of the vertical axis,
which is an order of magnitude too large.

\bibitem{ohno71} H. Ohno and M. Mekata, J. Phys. Soc. Jpn. {\bf 31}, 102 (1971).

\bibitem{kibey06} S. Kibey, J. B. Liu, M. J. Curtis, D. D. Johnson, and
H. Sehitoglu, Acta Mater. {\bf 54}, 2991 (2006).

\bibitem{dick09} A. Dick, T. Hickel, and J. Neugebauer, Steel Res. Int.
{\bf 80}, 603 (2009).

\bibitem{denteneer87} P. J. H. Denteneer and W. van Haeringen,
J. Phys. C Solid State Phys. {\bf 20}, L883 (1987).

\bibitem{martinez09} J. Martinez, S. M. Cotes, and J. Desimoni,
Phys. Status Solidi B {\bf 246}, 1366 (2009).

\bibitem{suzuki62} H. Suzuki, J. Phys. Soc. Jpn. 17, 322 (1962).

\bibitem{ueshima86} Y. Ueshima, S. Mizoguchi, T. Matsumiya,
and H. Kajioka, Metall. Trans. B {\bf 17}, 845 (1986).

\bibitem{li11} H. Li, L. Romaner, R. Pippan, and C. Ambrosch-Draxl
(in preparation).

\bibitem{vitos06} L. Vitos, P. A. Korzhavyi, and B. Johansson,
Phys. Rev. Lett. {\bf 96}, 117210 (2006).

\bibitem{gyorffy85} B. L. Gyorffy, A. J. Pindor, J. Staunton, G. M. Stocks,
and H. Winter, J. Phys. F Met. Phys. {\bf 15}, 1337 (1985).

\bibitem{abrikosov97} I. A. Abrikosov, S. I. Simak, B. Johansson, A. V. Ruban,
and H. L. Skriver, Phys. Rev. B {\bf 56}, 9319 (1997).

\bibitem{peil11} O.E. Peil, A.V. Ruban, and B. Johansson, submitted to PRB.

\bibitem{emto} L. Vitos, \textit{Computational Quantum Mechanics for
Materials Engineers} (Springer-Verlag, London, 2007).

\bibitem{lsgf_details} The local interaction zone in the LSGF-EMTO calculations
included the first two coordination shells in the case of the fcc structure and
the first three coordination shells in the case of the hcp one. Other details,
also  relevant for the EMTO-FCD calculations are as follows:
The partial waves have been expanded up to $l_{max}$ = 3.
The integration over the Brillouin zone has been performed using
k-mesh grids of 31x31x31 for fcc, 30x30x16 for hcp, and 30x30x8 for
dhcp structures.  The self-consistent electronic density calculations have
been done using the local spin-density approximation (LSDA)~\cite{perdew92},
while the total energy has been obtained in the generalized gradient
approximation (GGA)~\cite{perdew96}. 

\bibitem{perdew92} J. P. Perdew and Y. Wang, Phys. Rev B {\bf 45}, 13244
(1992).

\bibitem{perdew96} J. P. Perdew, K. Burke, and M. Ernzerhof, Phys. Rev. Lett.
{\bf 77}, 3865 (1996).

\bibitem{li02} C.-M. Li, F. Sommer, and E. J. Mittemeijer,
Mat. Sci. Engineering A {\bf 325}, 307 (2002).

\bibitem{a_300K} The experimental fcc data is available below $M_{s}$ because
the martensitic transformation is not complete, as found in earlier
experiments~\cite{umebayashi66}.

\bibitem{gebhardt10} T. Gebhardt, D. Music, B. Hallstedt, M. Ekholm,
I. A. Abrikosov, L. Vitos, and J. M. Schneider, J. Phys. Condens. Mat.
{\bf 22}, 295402 (2010).


\bibitem{moruzzi90} V.L. Moruzzi, Phys. Rev. B {\bf 41}, 6939 (1990).

\bibitem{moruzzi88} V.L. Moruzzi, J.F. Janak, and K. Schwarz,
Phys. Rev. B {\bf 37}, 790 (1988).

\bibitem{ruban07} A. V. Ruban, S. Khmelevskyi, P. Mohn, and
B. Johansson, Phys. Rev. B {\bf 75}, 054402 (2007).

\bibitem{cpa_details} The details of the EMTO-FCD calculations are the
same as in the LSGF-EMTO calculations~\cite{lsgf_details}, except from the fact
that the isomorphous model for the alloy has been used with the on-site
screening constant~\cite{ruban02} obtained from the corresponding
LSGF calculations. At $S_{WS}$ = 2.660~a.u. the $\alpha$ screening
constants~\cite{ruban02} are found to be 0.8143, 0.8120, and 0.8354 for
the fcc, hcp, and dhcp structures, respectively. 

\bibitem{ruban02} A. V. Ruban and H. L. Skriver, Phys. Rev. B {\bf 66},
024201 (2002); A. V. Ruban, S. I. Simak, P. A. Korzhavyi, and
H. L. Skriver, Phys. Rev. B {\bf 66}, 024202 (2002).

\bibitem{umebayashi66} H. Umebayashi, and Y. Ishikawa, J. Phys. Soc. Jpn.
{\bf 21}, 1281 (1966).


\end{thebibliography}
\end{document}